# Microstructural engineering by heat treatments of multi-principal element alloys via spinodal mediated phase transformation pathways

Shalini Roy Koneru[1], Kamal Kadirvel[1], Hamish Fraser and Yunzhi Wang[2]

Department of Materials Science and Engineering, The Ohio State University, Columbus, Ohio

**Abstract**

Nanoscale multi-phase microstructures observed in multi-principal element alloys (MPEAs) such as $AlMo_{0.5}NbTa_{0.5}TiZr$, $Al_{0.5}NbTa_{0.8}Ti_{1.5}V_{0.2}Zr$, TiZrNbTa, AlCoCrFeNi and $Fe_{15}Co_{15}Ni_{20}Mn_{20}Cu_{30}$ that exhibit promising mechanical or functional properties may have evolved through spinodal-mediated phase transformation pathways (PTPs). The microstructures in such MPEA systems could be further engineered for targeted applications by appropriately designing the alloy composition and heat-treatment schedule. In this study, we investigate systematically how different heat treatment schedules such as single-step isothermal aging, two-step isothermal aging and continuous cooling alter the interplay among the various factors associated with alloy composition, such as volume fraction of individual phases, lattice misfit and modulus mismatch between the co-existing phases. We have determined the degree to which these factors influence significantly the spinodal-mediated PTPs and the corresponding microstructures by use of high-throughput phase-field simulations. In particular, we demonstrate that the microstructural topology (i.e., which phase forms the continuous matrix and which phase forms discrete precipitates) in the same MPEA having an asymmetric miscibility gap could be inverted simply by a continuous cooling heat treatment. Further, we reveal a rich variety of novel hierarchical microstructures that could be designed using two-step isothermal aging heat treatments in MPEA systems with symmetric or asymmetric miscibility gaps. These simulation results may shed light on novel microstructure design and engineering for the above-mentioned MPEA systems.

---

[1] These authors contributed equally to this work.
[2] Corresponding author, Email: wang.363@osu.edu
Shalini Roy Koneru, Email: koneru.21@osu.edu
Kamalnath Kadirvel, Email: kadirvel.1@osu.edu
*Preprint submitted to Acta Materialia.

**Graphical Abstract**

# Microstructural engineering via heat treatments

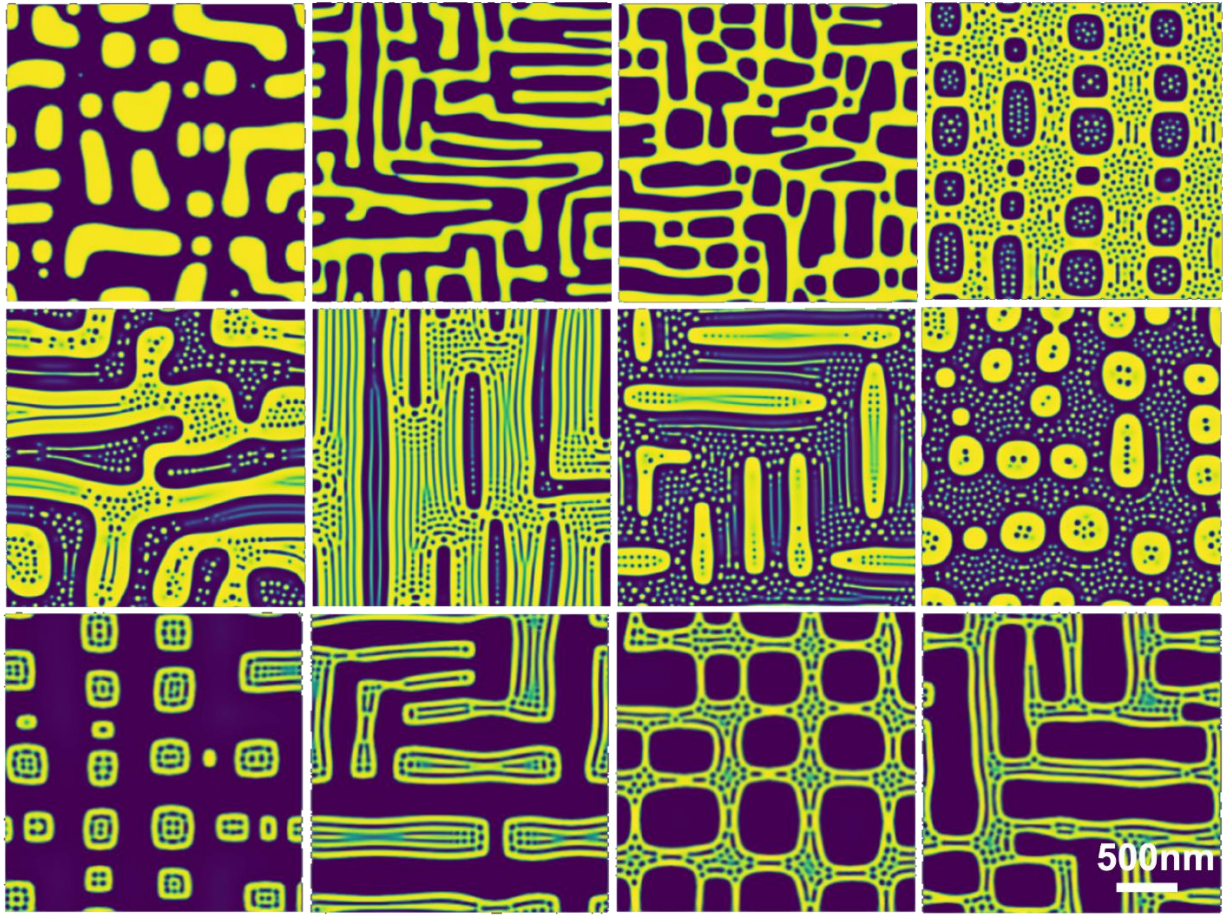

## 1. Introduction

Alloy design over the past decade has undergone a shift in alloying strategy from exploring only the corners of the compositional space to the centers of a multicomponent system to develop multi-principal element alloys (MPEAs) (also called high-entropy alloys (HEA)) [1–9]. The study of MPEAs suggests that the vast compositional design space offers such alloys many novel and sophisticated phase transformation pathways (PTPs) that could be used to engineer unique muti-phase microstructures with attractive balances of properties. Examples of MPEAs with such unique multi-phase microstructures include $Al_{19}Fe_{20}Co_{20}Ni_{41}$ [10] with $L1_2$+B2 lamellar microstructure, $Al_{0.3}CoFeNi$ [11] and $Al_{13}Fe_{29}Co_{29}Ni_{29}$ [12] with nano-lamellar microstructure consisting of alternate FCC+$L1_2$ and BCC+B2 lamellae, $Fe_{25}Co_{25}Ni_{25}Al_{10}Ti_{15}$ [13] that has FCC+B2 microstructure with hierarchical intragranular $L1_2$ nanoprecipitates in the FCC phase, HfZrTiNbTa with FCC/BCC/HCP mosaic microstructure, $Al_{0.5}NbTa_{0.8}Ti_{1.5}V_{0.2}Zr$ [14,15] and $AlMo_{0.5}NbTa_{0.5}TiZr$ [16–18] with nanoscale BCC+B2 inverted superalloy like microstructures and, TiZrNbTa [19] and $Fe_{15}Co_{15}Ni_{20}Mn_{20}Cu_{30}$ [20] with nanoscale BCC based and FCC based spinodal microstructures, respectively. Thus, the knowledge and high-throughput calculations of microstructural evolution in these multi-phase MPEAs and their dependence on the various degrees of freedom in the vast alloy design space including alloy composition, processing route, and heat treatment schedule are critical to engineering desired multi-phase microstructures for advanced applications. In this study, we focus on one of the frequently encountered among possible PTPs in MPEAs, namely spinodal assisted PTPs [21–26].

According to the literature, the decomposition of a single high-temperature disordered (BCC) phase into a mixture of ordered and disordered phases (BCC+B2) with the ordered (B2) phase as the matrix in $Al_{0.5}NbTa_{0.8}Ti_{1.5}V_{0.2}Zr$ MPEA could follow two possible spinodal-assisted phase transformation pathways (PTPs) [15,22,23]: i) congruent ordering of the disordered ($\beta$) phase followed by spinodal decomposition in the ordered $\beta'$ phase and then disordering of one of the ordered phases, i.e., $\beta \to \beta' \to \beta'_1 + \beta'_2 \to \beta' + \beta$; (ii) spinodal decomposition in the disordered BCC phase followed by ordering of one of the disordered phases, i.e., $\beta \to \beta_1 + \beta_2 \to \beta' + \beta$. These PTPs involving spinodal decomposition lead to the formation of nanoscale periodic two-phase microstructures with the spatial arrangement of the co-existing phases being dependent on the volume fractions of and lattice misfit and elastic moduli differences between the co-existing phases [22,23,27–30]. It is clear from these earlier works that under the assumption of homogeneous elastic moduli (i.e., the elastic constants of the co-existing phases are assumed to be the same), the phase with higher volume fraction would be the matrix which minimizes the interfacial area. This is different from precipitation via nucleation and growth where the precipitate phase, present as discrete particles, could have a volume fraction as high as 70% such as seen in some Ni-base superalloys [31]. For alloys with inhomogeneous moduli (i.e., the moduli of coexisting phases are different), the elastically more compliant phase (i.e., with lower elastic moduli) tend to wet the elastically stiffer phase and thus tend to be more connected and become the matrix, minimising the elastic energy, if its volume fraction is sufficiently large (although being significantly lower than 50%). For example, the BCC+B2 microstructures with B2 being the matrix are observed in $Al_{0.5}NbTa_{0.8}Ti_{1.5}V_{0.2}Zr$ [32] irrespective of the fact that the B2 volume fraction is as low as 30%. The elastic moduli mismatch (i.e., the B2 phase is more compliant than the BCC phase) has been proposed to be one of the potential factors for the observed microstructural topology. Note that the lattice mismatch between the coexisting phases dictates the elastic energy of the microstructure and thus the role played by the moduli mismatch in the optimization of the microstructural topology would be more significant for alloys with higher lattice mismatch. This implies that the design of this interplay between the volume fraction, lattice mismatch and moduli mismatch is vital in the design of microstructurally engineered MPEAs and it has already been studied in detail for alloys undergoing isothermal aging at a fixed temperature [22,27,29,30].

However, the volume fraction, lattice mismatch and moduli mismatch of co-existing phases change with temperature, and so the interplays among all these temperature-dependent factors is expected to change as temperature varies. This offers new opportunities to engineer microstructures simply via heat treatment schedule design without changing alloy composition. For example, instead of isothermal aging at a single temperature, continuous cooling and two-step aging could be employed to tailor the microstructures. In addition to temperature, cooling rate and aging time could also have significant impacts on microstructural development. For example, the matrix phase in the BCC+B2 microstructure of $Al_{0.5}NbTa_{0.8}Ti_{1.5}V_{0.2}Zr$ developed through spinodal assisted PTPs changes from B2 phase to BCC phase with just increasing the aging time and this microstructural change significantly modifies the mechanical properties of the alloy [15]. Thus, in this study, to investigate the effect of heat treatment schedule on the morphology (shape and spatial arrangement of the coexisting phases) and topology (which phase forms a continuous matrix and which phase forms discrete precipitates) of microstructures developed by spinodal-assisted phase transformation pathways, we simulate systematically using high-throughput phase-field simulations microstructural evolution under three relatively popular heat treatment schedules, i.e., single-step isothermal aging, two-step isothermal aging and continuous cooling for different alloy compositions.

Ni-based superalloys [33,34] with hierarchical microstructures (i.e., where additional $\gamma$ particles are present inside the $\gamma'$ phase particles) exhibits slower coarsening kinetics for the $\gamma'$ phase particles and much enhanced high temperature mechanical properties. Further, a recently developed $Fe_{25}Co_{25}Ni_{25}Al_{10}Ti_{15}$ HEA [13] with FCC+B2 microstructure that further has hierarchical intragranular $L1_2$ nanoprecipitates in the FCC phase exhibits a tensile yield strength of ~1.86GPa and a failure strain of ~5.2%. Also, the high entropy superalloys [35] with hierarchical microstructures exhibit better mechanical properties than those of the commercial single crystal superalloy CMSX-4. Thus, we explore the possibility of designing novel hierarchical microstructures with additional secondary phases present in the primary phases evolved through spinodal assisted PTPs by two-step aging [36] for possible unprecedented properties. Note that the microstructures simulated for the spinodal-assisted PTPs using phase-field simulations suggest that the spatial distribution of the co-existing phases in such systems is solely dependent on the spinodal decomposition step [22]. Thus, we investigate the effect of different heat treatment schedules on the microstructure evolution via spinodal decomposition or spinodal-assisted PTPs using phase-field simulations in simple prototype binary systems with symmetric or asymmetric miscibility gaps.

The remainder of the paper is organized as follows. We first present the phase-field model employed along with the prototype binary systems and the simulation parameters considered. Then, we compare and contrast microstructures simulated under different conditions (moduli mismatch and lattice misfit) for different alloys heat-treated using one-step isothermal aging compared with the use of continuous cooling for a system having an asymmetric miscibility gap. We quantify the connectivity and discreteness of these microstructures in an attempt to understand the effect of different heat treatment schedules on the interplay between the volume fraction, lattice misfit and moduli mismatch on microstructural evolution. Furthermore, we illustrate the rich variety of hierarchical microstructures that may be engineered in alloy systems with symmetric and asymmetric miscibility gaps via two-step isothermal aging heat treatments. Finally, we apply the knowledge derived from our simulations and use CALPHAD databases to suggest MPEAs with novel microstructures.

## 2. Phase-field model

In order to study the microstructural evolution, we considered a phase-field model with a single conserved order parameter, i.e., concentration $c(\vec{r}, t)$, that describes the solute concentration at a given point in the microstructure. The free energy functional of the system $F[c(\vec{r}, t)]$ is then formulated as

$$F[c(\vec{r}, t)] = \int \left[ f(c) + \frac{\kappa_c}{2} (\nabla c)^2 \right] dV + E^{el}[c] \tag{1}$$

where $f(c)$ is the local free energy, $\kappa_c$ is the gradient energy coefficient of concentration, and $E^{el}[c]$ is elastic energy functional. Note that the free energy of the system, $F[c(\vec{r}, t)]$, is assumed to be a functional of only the concentration field, $c(\vec{r}, t)$, in contrast to our earlier works [22,23] on microstructural evolution in alloys undergoing spinodal assisted PTPs where both concentration and structural order parameter fields, i.e., $c(\vec{r}, t)$ and $\eta(\vec{r}, t)$, were used. This is because these earlier works have already established that the final microstructures consisting of ordered + disordered phases in such alloy systems is dictated by the spinodal decomposition step rather than the congruent ordering or congruent disordering step.

### 2.1. Chemical energy of the system

In the present study, we considered two prototype binary systems, one with an asymmetric miscibility gap and the other with a symmetric miscibility gap. The local chemical free energy density $f(c)$ of the system as a function of temperature is given by

$$f(c) = X(1-X)[L_0 + L_1(1-2X) + L_2(1-2X)^2 + L_3(1-2X)^3] \\ + RT(X\ln(X) + (1-X)\ln(1-X)) \tag{2}$$

where, solute concentration $(X)$ could be written in terms of normalized concentration $(c)$ as $X = c * (X_y - X_b) + X_b$, $R$ is the gas constant and $T$ is the absolute temperature. Note that $X_b$ and $X_y$ represent the equilibrium concentrations of solute-lean and solute-rich phases at $0.5T_c$ for the symmetric prototype binary system and the normalization of concentration ensures that the equilibrium concentrations of the solute-lean and solute-rich phases at $0.5T_c$ are 0 and 1 respectively. Here, $T_c$ represents the critical spinodal decomposition temperature in the assumed miscibility gap. Similarly, for the system with an asymmetric miscibility gap, the normalized concentration is calculated based on the equilibrium concentrations $X_b$ and $X_y$ at $0.64T_c$. Also, we assumed a regular solution model with $L_0 = 16628$ J/mol and $L_1 = L_2 = L_3 = 0$ for the prototype binary system with a symmetric miscibility gap. For the prototype binary system with an asymmetric miscibility gap, we attempted to imitate the miscibility gap predicted for BCC phase in the $Al_5(TiZr)_x(NbTa)_{(95-x)}$ pseudobinary system using the RHEA database. To accomplish this, we utilized the PanOptimizer module in the Pandat$^{TM}$ software [37] to calculate the Redlich-Kister polynomial constants $(L_0, L_1, L_2, L_3)$ corresponding to the predicted asymmetric miscibility for $Al_5(TiZr)_x(NbTa)_{(95-x)}$ pseudobinary system, which yields $L_0 = 17333.9$ J/mol, $L_1 = 3832.3$ J/mol, $L_2 = 2835.5$ J/mol, $L_3 = 2703.1$ J/mol. The normalized phase diagrams i.e., the normalized temperature $\left(\frac{T}{T_c}\right)$ vs normalized concentration ($c$) depicting the miscibility gaps corresponding to the assumed free energy equations are shown in Figure 1(a) and 1(b), respectively.

## 2.2. Elastic energy of the system

The elastic energy functional of the system is given by

$$E^{el} = \int dr \left[\frac{1}{2} C_{ijkl}(\epsilon_{ij} - \epsilon_{ij}^T)(\epsilon_{kl} - \epsilon_{kl}^T)\right] \quad (3)$$

where $C_{ijkl}$ is the elastic modulus field, $\epsilon_{kl}^T$ is the stress-free transformation strain field and $\epsilon_{ij}$ is the total strain field. Both $C_{ijkl}$ and $\epsilon_{kl}^T$ are functions of the concentration field, $c(\vec{r}, t)$:

$$C_{ijkl}(c) = C_{ijkl}^b \times (1 - c) + C_{ijkl}^y \times (c) \quad (4)$$

$$\epsilon_{ij}^T = \epsilon^{oo} c \delta_{ij} \quad (5)$$

where $\epsilon^{oo} = \frac{1}{a}\frac{da}{dc}$ is the lattice misfit, $a$ is the lattice parameter, $C_{ijkl}^b$ and $C_{ijkl}^y$ are the moduli of the solute-lean and solute-rich phases, respectively. Here, the equilibrium compositions of the coexisting phases at various temperatures are different (as shown in Figure 1) and thus the total strain field $\epsilon_{ij}$ and elastic moduli field $C_{ijkl}$ would be temperature dependent owing to their composition dependence. However, the temperature dependences of the lattice misfit and moduli mismatch between the coexisting phases arising from their different temperature-dependence lattice parameters and elastic modulus are considered weak and negligible in this work in the temperature range considered. We assumed the modulus tensor to have cubic symmetry with Zener ratio $A_z = 2.0$, i.e., <001> directions have the lowest elastic modulus. The total strain field $\epsilon(\vec{r})$ is determined by the mechanical equilibrium equation: $\nabla . \sigma = 0$. For systems with homogeneous modulus (i.e., $C_{ijkl}^b = C_{ijkl}^y$), the elastic energy $E^{el}$ can be directly written as an explicit functional of the order parameters as derived by Khachaturyan et. al. [38,39] using the Green's function solution in the reciprocal space. Note that this closed form expression of $E^{el}$ greatly increases the numerical efficiency of the model. Further, the strain field is computed through an iterative procedure [40,41] for systems with inhomogeneous modulus where the modulus mismatch is quantified by a ratio $R_{elas} = C_{ijkl}^y/C_{ijkl}^b$. In this study, the microstructures are simulated for both homogeneous modulus case ($R_{elas} = 1.0$) and inhomogeneous modulus cases with $R_{elas} = 1.4$ and $R_{elas} = 0.6$ for the two phases.

## 2.3. Governing equation

The concentration field, $c(\vec{r}, t)$, is evolved with time using the Cahn-Hilliard [42] equation,

$$\frac{\partial c}{\partial t} = \nabla \left( M \nabla \left( \frac{\delta F}{\delta c} \right) \right) + \zeta_c \tag{6}$$

where $M$ is the chemical mobility, $\zeta_c$ is the Langevin-noise term. The semi-implicit spectral method [43] is utilized to solve Eq. (6). Note that the current study only concentrates on the evolution of microstructures through spinodal decomposition and thus the Langevin noise for concentration could be set to zero. However, it would take a long time for the accumulated numerical noises to act as fluctuations in concentration to initiate the spinodal decomposition. Thus, the Langevin noise for concentration with a small magnitude is used to accelerate the simulations.

## 2.4. Numerical Implementation

An in-house GPU code is used to numerically solve the phase-field equations. The non-dimensional model parameters utilized in the simulations are listed in Table 1. Note that the dimensionalized parameters could be calculated following the procedure described in ref [44]. We performed simulations in two dimensions with a system size of $512 \times 512$. Further, we considered identical Langevin noise in all the simulations to eradicate the statistical deviation in the spinodal microstructures between different alloys. Thus, the changes observed in the microstructures are solely due to the change in the model parameters.

## 2.5. Microstructure Quantification

Similar to our previous works, the simulated microstructures in this work are quantified using two parameters, connectivity parameter ($\theta$) and discreteness parameter ($\varphi$), respectively. A $3 \times 3$ supercell of the simulated microstructures is constructed and the number of solute-rich ($c > 0.5$) and solute-lean clusters ($c < 0.5$) are counted. We used the Hoshen-Kopelman algorithm [45] with von Neumann neighborhood to count the clusters. The parameters $\theta$ (matrix-parameter) and $\varphi$ (discreteness parameter) are defined as

$$\theta = \frac{N_B - N_Y}{N_B + N_Y}; \quad \varphi = |N_B - N_Y| \tag{7}$$

where $N_B$ and $N_Y$ are the number of solute-lean and solute-rich clusters respectively, which are represented by dark-blue and yellow colors, respectively, in the microstructure plots (see, e.g., Figure 2). Here, note that the microstructures with higher discreteness have higher $\varphi$ values. The microstructure with an isolated solute lean particle in a solute-rich matrix correspond to $\theta = 0.8$, whereas the microstructure with an isolated solute rich particle in a solute-lean matrix has $\theta = -0.8$. Thus, $\theta \geq 0.8$ and $\theta \leq -0.8$ indicate microstructures with solute-rich phase (yellow) and solute-lean phase (dark-blue) matrix, respectively. Further, $-0.8 < \theta < 0.8$ suggests bi-continuous microstructures.

## 3. Results

It is obvious from the prototype binary phase diagrams (shown in Figure 1) that the volume fractions and compositions of the equilibrium phases are a function of temperature. This implies that the interplay among the volume fraction of, lattice misfit and modulus mismatch between, the coexisting phases during the microstructure evolution would also depend on the heat treatment schedule apart from alloy composition. To explore this in detail, we simulate the microstructure evolution under three relatively popular heat treatment schedules i.e., single-step isothermal aging, two-step isothermal aging and continuous cooling for different alloy compositions. These heat treatment schedules (shown in Figure 1(c)-(e)) as normalized temperature vs simulation time plots) are described below.

*Asymmetric miscibility gap prototype system:*

i. Single-step isothermal aging (HT1): The isothermal aging temperature is selected here to ensure that all the alloy compositions explored are well inside the spinodal curve. Furthermore, we intend to design the continuous cooling and two-step isothermal aging heat treatments with this selected temperature as the final temperature and, thus, the selected temperature should not be too high to accommodate the design of both these heat treatments. Thus, the solutionized alloys are aged at temperature $0.64T_c$ in HT1. The alloy selection is explained in the next section.

ii. Continuous cooling (HT2): Continuous cooling is the most practical heat treatment schedule employed in industry. We hypothesize that microstructure evolution during continuous cooling could be quite different from that during isothermal aging owing to the temperature-dependance of the interplay among competing factors during the microstructural evolution. To test this hypothesis, we simulate microstructure development in the solutionized alloy that is continuously cooled from $0.91T_c$ to $0.64T_c$. Note that the continuous decrease in temperature is simulated through discrete isothermal steps in the phase-field simulations as shown in Figure 1(d).

iii. Two-step isothermal aging (HT3): We aim to achieve a range of hierarchical microstructures through the design of two-step isothermal aging heat treatments. Thus, the aging temperatures for the two-steps are selected in such a way that at least one of the coexisting phases in the microstructure generated during the first aging step are inside the spinodal curve at the second aging temperature. Thus, the solutionized alloy is initially aged at $0.87T_c$, then cooled to $0.64T_c$ and further aged at that temperature.

*Symmetric miscibility gap prototype system:*

Note that both the volume fractions and compositions of the equilibrium phases change dramatically in the prototype binary system with an asymmetric miscibility gap (Figure 1(a)). Thus, we restrict our investigations of single-step isothermal aging vs continuous cooling to the system with an asymmetric miscibility gap as it would demonstrate the temperature effect the best, as will be further discussed in detail in Section 4.1. However, for the two-step aging heat treatment, the symmetric nature of the miscibility gap would let both the equilibrium phases corresponding to the first aging step to be inside the spinodal during the second aging step and this may lead to interesting hierarchical microstructures. Thus, we simulate microstructure evolution during two-step aging for the system with a symmetric miscibility gap (Figure 1(b)).

   i.  Two-step isothermal aging (HT3): The solutionized alloy is initially aged at $0.9T_c$ and then cooled to $0.5T_c$ and further aged at this temperature.

## 3.1. Alloys investigated

The choice of alloy composition could affect the role played by the heat treatment schedule in the interplay among the competing factors influencing the microstructural evolution as it determines the volume fraction of the coexisting phases at equilibrium. Thus, we selected three alloy compositions within the spinodal aimed for a comprehensive understanding of the interplay between the effects of heat treatments and alloy composition on the microstructure morphology and topology.

*Asymmetric miscibility gap prototype system:*

Three alloy compositions, *AC1, AC2* and *AC3,* have been chosen for the system with an asymmetric miscibility gap, as indicated by the three vertical dashed lines in Figure 1(a). They have significantly different variations in volume fractions of the coexisting phases as a function of temperature. *AC1* has the critical composition, i.e., at the position of the peak of the miscibility gap, $c = 0.22$, where the decomposition will start with equal volume fractions of the two-coexisting phases at temperatures just below the critical point, but drastically different volume fractions at lower temperatures. *AC2* ($c = 0.32$) and *AC3* ($c = 0.39$) compositions are selected as the solute-rich phase evolves from being major phase to minor phase with decreasing temperature in these alloys. Further, selection of both *AC2* and *AC3* enables us to study the effect of continuous cooling vs single-step isothermal aging on microstructure evolution in alloys with different solute-rich phase fraction at $0.64T_c$ (HT1 temperature). We simulate microstructural evolutions in these alloys during single-step isothermal aging (HT1) and continuous cooling (HT2) under different conditions of lattice misfit and modulus mismatch between the coexisting phases to document the effect of heat treatment schedule on the interplay between the volume fraction of, lattice misfit and modulus mismatch of coexisting phases. Note that these compositions are normalized compositions (as explained in Section 2.1) to ensure that the equilibrium compositions at $0.64T_c$ are 0 and 1. Thus, the normalized composition ($c$) also represents the volume fraction of the solute-rich phase at $0.64T_c$ for the given alloys.

Further, we simulated microstructural evolutions in *AC1* ($c = 0.22$) and *AC2* ($c = 0.32$) during two-step aging treatments under different conditions of lattice misfit and modulus mismatch between the coexisting phases with a hypothesis to generate various hierarchical microstructures.

*Symmetric miscibility gap prototype system:*

For the system with a symmetric miscibility gap, we simulate microstructural evolutions also in three alloys that have different volume-fractions of the solute rich phase, i.e., *SC1* ($c = 0.4$), *SC2* ($c = 0.5$) and *SC3* ($c = 0.6$) as indicated by the three dashed lines in Figure 1(b). Here, the normalized composition ($c$) in this case represents the volume-fraction of the solute-rich phase at $0.5T_c$. Note that the alloy compositions are selected such that *SC1* and *SC3* have solute-rich phases as minor and major phase respectively, whereas *SC2* has equal volume fraction of solute-rich and solute-lean phases.

*Elastic properties and lattice misfit of the prototype systems:*

As explained earlier, we simulate the microstructures under different elasticity conditions (modulus mismatch), such as i) *Homogeneous modulus*: moduli of the equilibrium phases are independent of composition, i.e., $R_{elas} = 1$ ($R_{elas}$ defined in Section 2.2); ii) *Hard solute-rich phase*: the solute rich phase has a higher modulus, i.e., $R_{elas} = 1.4$; iii) *Soft solute-rich phase*: the solute rich phase has a lower modulus, i.e., $R_{elas} = 0.6$, to illustrate the effect of heat treatment schedule on microstructures in alloys with different modulus mismatch values. Also, we explored the effect of different lattice misfit values on the microstructural evolution by simulating the microstructures for asymmetric system with $\epsilon^{oo} = 0.1\%$ and $\epsilon^{oo} = 0.15\%$. For the symmetric miscibility gap case, the microstructures are simulated under lattice misfits of $\epsilon^{oo} = 0.05\%$ and $\epsilon^{oo} = 0.1\%$.

## 3.2. Microstructures developed in alloys processed using isothermal aging vs continuous cooling

### 3.2.1. Alloys having different phase fractions under homogeneous modulus condition

The microstructures simulated under the homogeneous modulus condition for alloys of three different compositions in the asymmetric binary system (Figure 1(a)) processed through two different heat treatments HT1 (isothermal aging) and HT2 (continuous cooling) are shown in Figure 2. Note that a dilatational stress-free transformation strain, $\epsilon^{oo} = 0.1\%$, is assumed in these simulations. The effect of different heat treatments on the spatial arrangement of the solute-rich (yellow) and solute-lean (dark-blue) phases in different alloys is summarized below.

For the alloy with $c = 0.22$ (i.e., *AC1* with the volume fraction of solute-rich phase $v_f = 0.22$ at equilibrium), the simulated microstructures for different heat treatment schedules exhibit a solute-lean matrix with discrete solute-rich phase particles owing to the large phase volume fraction difference. However, a certain fraction of the solute-rich phase is present as elongated particles in the microstructure corresponding to HT2 as shown in Figure 2. For the alloy with $c = 0.32$ (i.e., *AC2* with $v_f = 0.32$), the microstructures obtained under the two heat treatment conditions are quite different for the same alloy. The microstructure corresponding to HT1 has discrete solute-rich phase particles present in the solute-lean matrix, whereas the microstructure corresponding to HT2 consists of an highly interconnected solute-rich phase in an almost bi-continuous microstructure. The detailed quantifications of the continuity and discreteness of the microstructures are provided in the next section. For the alloy with $c = 0.39$ (i.e., *AC3* with $v_f = 0.39$), the solute-rich phase exits as discrete particles in the solute-lean matrix phase in the microstructure corresponding to HT1, whereas discrete solute-lean phase particles are present in the solute-rich matrix phase in the microstructure corresponding to HT2, i.e., *the matrix phase has been inverted in the same alloy processed under these two different heat treatment conditions*. The linear intercepts of interfaces are also significantly different in these two microstructures.

As specified earlier, it is generally expected that the phase with higher volume fraction would be the matrix phase (as observed in alloys processed through HT1) under the homogeneous modulus condition. According to the existing studies [22,23,27–30,46], the topology could be inverted, i.e., the phase with lower volume fraction could be the matrix phase if its modulus is significantly lower than that of the other phase. Here, however, we observe that *the phase with lower volume fraction ($v_f = 0.39$) is the matrix phase in AC3 processed through HT2 even under the homogeneous modulus condition*. To further understand the phase inversion in detail in such alloys, microstructural evolutions in *AC3* under different heat treatment conditions are shown in Figure 3. It is readily seen from the microstructures of HT1 (single-step isothermal aging at $0.64T_c$) that the yellow phase (solute-rich) becomes the discrete phase from the early stages. However, the early microstructures formed under HT2 (continuous cooling) have almost a bi-continuous topology, but with a few discrete solute-lean (dark-blue) particles, i.e., the solute-rich phase is actually the matrix phase at early stages, and this leads to the final microstructure having the yellow phase as the matrix. According to Figure 1(a), *AC3* enters the spinodal region at approximately $0.88T_c$. Because of the asymmetry of the miscibility gap, the solute-rich phase (yellow) has a much higher volume fraction than that of the solute-lean phase at this temperature. Thus, the solute-rich phase emerges as the matrix phase in the early stages of spinodal decomposition at temperatures greater than $0.64T_c$. *This implies that in systems with an asymmetric miscibility gap, different heat treatment schedules could be utilized to design microstructures with different topologies (i.e., continuous vs. discrete phases) for certain alloys*.

### 3.2.2. Alloys having different phase fractions and lattice misfit under inhomogeneous modulus condition

It is clear from the above simulation results that continuous cooling could be utilized for systems with asymmetric miscibility gaps to render the lower volume fraction phase ($v_f$ as low as 0.39) to be the matrix phase under homogeneous modulus condition. Here, we further explore the interplay between the effect of processing through different heat treatments on the microstructures of alloys having different phase fractions and lattice misfit under *inhomogeneous modulus condition*. The microstructures of the same alloys *AC1, AC2* and *AC3* shown in Figure 1(a) simulated under different combinations of parameters such as heat treatment schedule, elasticity conditions and lattice misfit are shown in Figure 4, Figure 5 and Figure 6, respectively. The quantification parameters for connectivity ($\theta$) and discreteness ($\varphi$) of the simulated microstructures are shown in Figure 7.

For *AC1* ($c = 0.22$) *with* $\epsilon^{oo} = 0.1\%$ (left panel of Figure 4), all the microstructures simulated under different conditions have discrete solute-rich (yellow) phase particles in the solute-lean matrix. Thus, all the microstructures with solute-lean matrix exhibit connectivity parameter $\theta < -0.8$ as shown in Figure 7(a) (see Eq. (7) for the definition of θ). Similar to the homogeneous modulus condition, we observe a small fraction of elongated particles in microstructures obtained under HT2. This leads to a lower discreteness $\varphi$ value in HT2 microstructures as shown in Figure 7(g).

For *AC1* ($c = 0.22$) *with $\epsilon^{oo} = 0.15\%$* (right panel of Figure 4), it is interesting to observe mostly elongated yellow particles along elastically soft directions in most of the microstructures simulated under different conditions except in the isothermally aged (HT1) alloy microstructure simulated under the "hard" yellow phase condition. We observe nearly rounded-square discrete yellow particles in this alloy owing to the higher modulus of the yellow phase and, thus, it exhibits the highest discreteness $\varphi$ in comparison with the other microstructures for *AC1* as shown in Figure 7(h). Under HT2, considerably larger fractions of the higher modulus yellow phase appear as elongated particles as compared with the cases under HT1. This could be attributed to the start of spinodal decomposition at temperatures higher than $0.64T_c$ in HT2, which will be discussed further in detail in the discussion section. We further notice that the elongated yellow phase particles in *AC1* simulated under HT2 and soft yellow phase condition are interconnected to form maze-like microstructures (the bottom right figure in Figure 4). All the microstructures exhibit $\theta < -0.8$ (shown in Figure 7(b)), consistent with visual observations. Similar to the trend observed in microstructures with $\epsilon^{oo} = 0.1\%$, the microstructures of *AC1* processed through HT1 have the highest discreteness $\varphi$.

In the case of *AC2* ($c = 0.32$) with $\epsilon^{oo} = 0.1\%$ (the left panel in Figure 5), the microstructures simulated for the alloys processed under HT1 exhibit discrete yellow phase particles irrespective of the different elasticity conditions, with $\theta < -0.8$ (triangles in Figure 7(c)). The majority of the yellow phase particles simulated under soft yellow phase appear as elongated particles in contrast to the nearly equiaxed particles observed in microstructures simulated under homogeneous modulus and hard yellow phase conditions. In the case of continuous cooling (HT2), the microstructure obtained under soft yellow phase condition has $\theta > 0.8$ - with solute-lean (dark-blue) phase discrete particles embedded in the continuous yellow matrix, the microstructure obtained under a hard yellow phase condition has $\theta < -0.8$ - with yellow discrete particles embedded in the continuous dark-blue matrix, and the microstructure simulated under the homogeneous modulus condition has a nearly bi-continuous topology with $\theta = -0.23$. *It is interesting to learn that the soft yellow phase with volume fraction as low as 30% could become the matrix, i.e., the topology of the microstructure could be completely inverted by continuous cooling vs. isothermal aging*. Also, it is evident from Figure 7(i) that the microstructure corresponding to the iso-thermally aged alloy has the highest discreteness $\varphi$ in comparison with that of the continuously cooled alloys.

For *AC2* ($c = 0.32$, Figure 5) with $\epsilon^{oo} = 0.15\%$, the trends observed in the microstructure topology (i.e., the continuous vs discrete phases shown in the right panel of Figure 5 as well as the connectivity parameter shown in Figure 7(d)) are the same as those observed in the microstructures simulated for *AC2* with $\epsilon^{oo} = 0.1\%$ under different conditions. Note that the discreteness of the microstructures of isothermally aged *AC2* with $\epsilon^{oo} = 0.15\%$ (shown in Figure 7(j)) simulated under different conditions is lower in comparison with those observed in the alloys with $\epsilon^{oo} = 0.1\%$. However, the discreteness of the microstructures of continuously cooled *AC2* with $\epsilon^{oo} =$ 0.15% simulated under different modulus mismatch conditions is relatively higher in comparison with the alloys with $\epsilon^{oo} = 0.1\%$ (shown as red bars in Figure 7 (i) and (j)).

For *AC3* ($c = 0.39$, Figure 6) with $\epsilon^{oo} = 0.1\%$, the microstructures corresponding to HT1 simulated under different elasticity conditions have discrete yellow particles with $\theta < -0.8$ (Figure 7(e)). The microstructures corresponding to HT2 simulated under the homogeneous modulus and soft yellow phase conditions exhibit the opposite, i.e., discrete dark-blue phase particles embedded in a continuous yellow phase matrix with $\theta > 0.8$. The discreteness in the softer yellow phase case is higher as seen in Figure 7(k). In the case of harder yellow phase, we observe a microstructure with the dark-blue phase being the matrix (with $\theta < -0.8$). The discreteness of the microstructures in alloys processed under HT1 is higher than their continuously cooled alloy counterparts in cases with homogeneous modulus and hard solute-rich phase, whereas the discreteness of the continuously cooled *AC3* microstructure simulated under soft yellow phase is higher than its isothermally aged counterpart.

In the case of *AC3* ($c = 0.39$) with $\epsilon^{oo} = 0.15\%$ (right panel of Figure 6) processed through HT1, we observe discrete yellow particles embedded in a continuous dark-blue matrix in the microstructures simulated under homogeneous modulus and hard yellow phase conditions with $\theta < -0.8$ (shown in Figure 7(f)), whereas the microstructure simulated under the soft yellow-phase condition exhibits $\theta = 0.72$ with relatively larger number of dark-blue particles. This implies that the elastic strain energy under the assumption of $\epsilon^{oo} = 0.15\%$ in the simulations is sufficient to counteract the volume fraction effect and render the microstructure with soft yellow phase fraction as low as 0.39 to have $\theta$ as high as 0.72 instead of $-0.8$ observed under homogeneous modulus condition in isothermally aged alloys. Furthermore, similar to the case with $\epsilon^{oo} = 0.1\%$, discrete dark-blue particles embedded in a continuous yellow matrix phase with $\theta > 0.8$ is observed in continuously cooled (HT2) alloy microstructures simulated under soft yellow phase condition, whereas the microstructure simulated under hard yellow phase condition exhibits a continuous dark-blue matrix phase with $\theta < -0.8$. However, the microstructure simulated under homogeneous modulus condition is bi-continuous with $\theta = -0.02$ for the continuously cooled alloy. This is in contrast to the discrete dark-blue phase particle microstructures observed in simulations under homogeneous modulus condition for the case with $\epsilon^{oo} = 0.1\%$. The trends observed in the discreteness of the microstructures (Figure 7(l)) are almost similar to those observed for the case with $\epsilon^{oo} = 0.1\%$ (Figure 7(k)), except that the continuously cooled microstructure simulated under homogeneous modulus condition has lower discreteness in comparison with that simulated under hard yellow phase condition owing to its bi-continuous topology with $\theta = -0.02$.

In all the cases of *AC2* and *AC3*, we noted that the interphase interface density is significantly higher in microstructures developed under continuous cooling (HT2) as compared to those obtained under isothermal aging (HT1).

## 3.3. Design of hierarchical microstructures by two-step isothermal aging

### 3.3.1. *Alloy system with a symmetrical miscibility gap*

To demonstrate the possibility of using two-step isothermal aging heat treatments to engineer hierarchical microstructures, we first simulate microstructural evolution in an alloy with $c = 0.6$ (i.e., *SC3* with the volume fraction of solute-rich phase $v_f = 0.6$ at $0.5T_c$) in the prototype binary system with a symmetric miscibility gap (Figure 1(c)) under zero lattice misfit condition. The simulation result is shown in Figure 8. The two-phase microstructure developed is certainly unique, with the second generation of spinodal decomposition at the lower aging temperature generating much finer two-phase microstructures within each phase domains generated by the first-generation of spinodal decomposition at the higher aging temperature (Figure 8(d)-(e)). The microstructures captured at the end of the first isothermal aging step (Figure 8(b)) and during the early stages of the second isothermal aging step (Figure 8(c)) demonstrate clearly the formation process of such a hierarchical microstructure. The solute-rich and solute-lean domains formed at the end of the first isothermal aging step re-enter the spinodal again at the second isothermal aging step, and the spinodal decomposition in such a heterogeneous system starts from the existing inter-phase interfaces and propagates into both the solute-rich and solute-lean primary phase domains. Note that the finer discrete particles (secondary) in the yellow matrix (Figure 8(e)) are an outcome of superposition of the concentric concentration waves generated from different sources (i.e., the primary particles). Thus, the spatial arrangement of the coexisting phases in the final hierarchical microstructures developed is highly dependent on the spatial arrangement of the phases in the microstructure at the end of the first isothermal aging heat treatment step similar to the microstructural evolution predicted in polymer blends [36] under two-step quenching heat treatments. Interestingly, the microstructural evolution shown in Figure 8(d)-(e) further suggests that a longer aging time during the second isothermal aging step leads to coarsening of the second generation of particles and thus reduces the scale difference between the two microstructure hierarchies. Thus, the selection of appropriate aging times in HT3 is also vital in the design of alloys with hierarchical microstructures.

To further explore the variety of hierarchical microstructures that could be engineered in systems with a symmetric miscibility gap, we simulate microstructural evolution during two-step isothermally aging (Figure 1(e)) of alloys with different volume fractions of, lattice misfit, and modulus mismatch between the coexisting phases. The simulation results are shown in Figure 9. It is readily seen that the interplay among these factors dictates the morphology (shape and spatial arrangement of phase domains) and topology (which phase forms discrete particles and which phase forms a continuous matrix) of the microstructures at the end of the first isothermal aging heat treatment, which form the template for the development of the second-generation of the hierarchical precipitate microstructures. When the lattice misfit is relatively small, e.g., $\epsilon^{oo} = 0.05\%$, the volume fraction effect dominates over the modulus mismatch effect in the development of the primary (first generation) two-phase domain structures (i.e., phase separation at the first aging step) and, thus, we observe solute-rich primary phase (yellow) discrete particles, bi-continuous, and discrete particles of the solute-lean primary phase (dark-blue) in alloys with $c = 0.4, 0.5$ and $0.6$ respectively (Figure 9, left panel).

In alloys with $\epsilon^{oo} = 0.1\%$, we observe discrete particles of solute-rich and solute-lean primary phases in alloys with $c = 0.4$ and $0.6$, respectively (Figure 9, the right panel, left and right columns, respectively). The discrete particles have elongated shapes (along the elastically soft directions) except in the alloy with $c = 0.6$ simulated under soft yellow phase condition where the solute-lean (dark-blue) discrete particles have cuboidal shapes, similar to the hierarchical microstructures observed in superalloys [33,34] and high entropy superalloys [35]. For the alloys with $c = 0.5$, we observe microstructures with bi-continuous primary phases in alloys with homogeneous modulus or hard yellow phase, whereas a microstructure with discrete particles of the primary dark-blue phase is observed in the alloy with the soft yellow phase. The co-existing phases are highly aligned along elastically soft directions in all 9 cases considered.

The formation of the ultrafine second-generation (secondary) solute-rich and solute-lean particles within the first-generation (primary) solute-lean and solute-rich domains follows a similar process as described above for the case with zero lattice misfit.

### 3.3.2. Alloy system with an asymmetrical miscibility gap

The microstructures developed during two-step isothermally aging (Figure 1(e)) in alloys with an asymmetric miscibility gap (Figure 1(a)) and having different volume fractions of, lattice misfit and modulus mismatch between the coexisting phases are shown in Figure 10. In these alloys only one of the primary coexisting phases, i.e., the solute-rich (yellow) phase, re-enters the spinodal during the second isothermal aging step and, thus, we observe a different type of hierarchical microstructure, with the formation of a fine two-phase mixture only in one of the primary phase domains, i.e., the solute-rich (yellow) domains.

After the first isothermal aging step at $0.87T_c$, we observe microstructures with discrete primary solute-rich (yellow) phase particles in alloys with $c = 0.22$ and $\epsilon^{oo} = 0.1\%$ irrespective of the modulus mismatch between the coexisting phases. However, elongated shapes aligned along the elastically soft directions are observed in microstructures obtained under homogeneous modulus and, in particular, soft yellow phase conditions, whereas the microstructures simulated under hard solute-rich condition has more or less equiaxed discrete solute-rich particles. Furthermore, our simulations suggest that the alloys with $c = 0.22$ and $\epsilon^{oo} = 0.15\%$ under the hard solute-rich phase and homogeneous modulus conditions exhibit microstructures with discrete solute-rich particles (of different degrees of discreteness), whereas the microstructure simulated under soft solute-rich phase condition exhibits discrete solute-lean particles. For the case of alloys with $c = 0.32$ (with either $\epsilon^{oo} = 0.1\%$ or $\epsilon^{oo} = 0.15\%$), we observe microstructures with discrete primary solute-lean phase particles in simulations under homogeneous modulus and soft solute-rich phase conditions, whereas microstructures with discrete primary solute-rich phase particles are observed in simulations under hard solute-rich phase conditions.

These two-phase microstructures formed at the end of the first isothermal aging serve as the templates for the formation of hierarchical microstructures with varied hierarchical architecture during the second isothermal aging step, as shown in Figure 10. Thus, hierarchical microstructures with varied architectures ranging from having a distribution of secondary precipitates only in the matrix phase to presence of secondary precipitates only in discrete particles could be engineered in systems with asymmetric miscibility gap. The hierarchical architecture in microstructures of alloys with $c = 0.32$ and soft solute-rich phase is similar to the spatial distribution of phases in superalloy with secondary $\gamma'$ precipitates [47].

## 4. Discussion

Our simulations demonstrate that the microstructural topology for certain alloys like *AC2* with a soft solute rich phase could vary from having a solute-lean matrix, a solute-rich matrix, or hierarchical microstructures, and the microstructural morphology could vary from having nearly rounded square discrete particles to elongated discrete particles with just varying the heat treatment from single-step isothermal aging to continuous cooling to two-step aging. The role played by different parameters, such as an asymmetricy miscibility gap, volume fraction, lattice misfit, and modulus mismatch between the coexisting phases, on this effect of the heat treatment schedule is discussed in detail below.

### 4.1. Microstructural topology inversion via continuous cooling

As mentioned earlier, the microstructural topology inversion (or phase inversion) observed in *AC3* with change in heat treatment schedule from isothermal aging to continuous cooling shown in Figure 3 is a consequence of the initiation of spinodal decomposition at temperatures higher than $0.64T_c$ during continuous cooling. To further understand the role of temperature, we examined the variation in solute-rich phase fraction $(v_f)$ with normalized temperature $\left(\frac{T}{T_c}\right)$ and the results are shown in Figure 11(a). In all the alloys except for *AC1*, *the major phase at equilibrium changes from a solute-rich to a solute-lean phase with decreasing temperature within the studied temperature range in the system with an asymmetric miscibility gap* (Figure 1(a)). This implies that the solute-rich (yellow) phase could be the matrix phase in alloys spinodally decomposed at higher temperatures with $v_f > 0.5$ under homogeneous modulus condition owing to its higher volume fraction. Thus, the early microstructures of *AC3* (Figure 3) evolved through continuous cooling from $0.91T_c$ have yellow matrix phase and this is retained in the final microstructures observed at the simulation time $t = 28000$. Note that this variation in the solute-rich phase fraction from being the major phase to being the minor phase, i.e., from $(v_f > 0.5)$ to $(v_f < 0.5)$ upon cooling is because of the asymmetry in the miscibility gap and does not exist in systems with a symmetric miscibility gap. Thus, *the asymmetry in the miscibility gap could be utilized to design alloys with desired microstructural topology by controlling the heat treatment process*.

### 4.2. Interplay between the effects of alloy composition and heat treatment schedule

As shown in Figure 11(a), the alloy composition controls the variation in volume fractions of the co-existing phases with temperature in an asymmetrical miscibility gap, and this could alter the effect of single-step isothermal aging vs continuous cooling on the microstructure topology. Thus, we compare the microstructures developed under *homogeneous moduli condition* ($R_{elas} = 1$) in this section and discuss the effect of the interplay between the alloy composition and the heat treatment schedule on the microstructure topology. The lower solute-rich phase fraction at the final aging temperature $0.64T_c$ suggests that alloys *AC1, AC2* and *AC3* exhibit microstructures with discrete solute-rich phase particles in systems with $R_{elas} = 1$ [22,27,29,30]. During continuous cooling, however, the volume fractions of the solute-rich and solute-lean phases could deviate significantly from those at the final aging temperature because of the asymmetry of the miscibility gap as shown in Figure 11(a). As a consequence, the effect of continuous cooling (HT2) for *AC2* and *AC3* counteracts the equilibrium volume fraction effect, and the alloys exhibit bi-continuous microstructures or microstructures with a solute-rich matrix (Figure 7). For the alloy composition corresponding to the peak of the miscibility gap, i.e., *AC1*, the dramatic change of the solute-rich phase from being major phase to minor phase with decreasing temperature would be absent (as shown in Figure 11(a)) and thus we observe microstructures with discrete solute-rich particles in both isothermally aged and continuously cooled samples of *AC1*.

The discreteness ($\varphi$) of the simulated microstructures of alloys *AC1, AC2* and *AC3* (with $R_{elas} = 1$) processed through HT1, shown in Figure 7(g)-(l), suggest that $\varphi$ is the highest for *AC1* irrespective of the assumed lattice misfit ($\epsilon^{oo}$) owing to its lowest solute-rich phase fraction at $0.64T_c$. As explained earlier, the initiation of spinodal decomposition at temperatures higher than $0.64T_c$ led to higher connectivity of the solute-rich phase and thus the microstructures of alloys *AC1, AC2* and *AC3* processed through continuous cooling has lower discreteness in comparison to their isothermally aged counterparts under homogeneous modulus condition. Further, this initiation of spinodal decomposition at temperatures higher than $0.64T_c$ during continuous cooling also led to higher interphase interface density in microstructures of alloys *AC2* and *AC3* processed through continuous cooling.

## 4.3. Interplay between the effects of alloy composition, modulus mismatch ($R_{elas}$) and heat treatment schedule

We observe microstructures with a solute-rich phase matrix (i.e., with $\theta > 0.8$ as shown in Figure 7) in continuously cooled alloys *AC2* and *AC3* with $R_{elas} < 1$ and microstructures with solute-lean phase matrix (with $\theta < -0.8$) in all continuously cooled alloys with $R_{elas} > 1$. This could be attributed to the interplay between the temperature dependent phase fraction and the elastic modulus mismatch. While the relatively high volume fractions of the solute-rich phase at early stage of decomposition taking place at higher temperatures during continuous cooling of these alloys (see the asymmetric miscibility gap shown in Figure 1(a)) always favors the solute-rich phase as the continuous matrix phase, the $R_{elas} > 1$ (i.e., hard solute-rich phase) condition counteracts this temperature dependent volume fraction effect and the $R_{elas} < 1$ (i.e., soft solute-rich phase) condition enhance this effect, as the elastically more compliant phase tends to wet the elastically stiffer phase. Note that the solute-rich phase volume fraction in *AC1* (having composition located at the peak of the miscibility gap) is too low at all temperatures to attain microstructures with the solute-rich phase as the matrix phase irrespective of the condition of continuous cooling with $R_{elas} < 1$.

Our simulations suggest that out of the 36 alloys considered (i.e., *AC1, AC2* and *AC3* under different conditions) (Figure 7), microstructures with the solute-rich phase as the matrix phase have the highest discreteness in continuously cooled *AC3* with $v_f = 0.39$, $R_{elas} < 1$ and $\epsilon^{oo} = 0.15\%$. This is because both the modulus mismatch effect and the temperature-dependent phase fraction effect through continuous cooling enhance the discreteness of the solute-lean phases. In contrast, microstructures having the solute-lean phase as the matrix phase reach the highest discreteness in isothermally aged *AC1* with $v_f = 0.22$, $R_{elas} > 1$ and $\epsilon^{oo} = 0.15\%$, owing to the lowest volume fraction of a stiffer solute-rich phase at the isothermal aging temperature. A bi-continuous microstructure with almost zero discreteness is observed in continuously cooled *AC3* with $v_f = 0.39$, $R_{elas} = 1$ and $\epsilon^{oo} = 0.15\%$. Usually, we expect such a bi-continuous microstructure with almost zero discreteness in isothermally aged alloys with $v_f = 0.5$ and $R_{elas} = 1$. However, exploitation of heat treatment schedule as a design parameter allows us to achieve such a microstructure in alloys with $v_f \ll 0.5$ under the same elasticity condition (i.e., $R_{elas} = 1$).

### 4.4. Effect of lattice misfit

The influence of modulus mismatch ($R_{elas} \neq 1$) on microstructure topology will be higher in systems with higher lattice misfit ($\epsilon^{oo}$) owing to the strong dependence of elastic strain energy on $\epsilon^{oo}$, i.e., the elastically stiffer phase will tend to be more discrete and the compliant phase will be more connected at higher $\epsilon^{oo}$. Thus, we observe a highly connected solute-rich phase (Figure 4) in isothermally aged *AC1* with $R_{elas} < 1$ and $\epsilon^{oo} = 0.15\%$ in contrast to its counterpart with $\epsilon^{oo} = 0.1\%$. Similarly, we observe a microstructure with a highly connected soft solute-rich phase (Figure 6) and $\theta = 0.72$ (Figure 7(f)) in isothermally aged *AC3* with $v_f = 0.39$, $R_{elas} < 1$ and $\epsilon^{oo} = 0.15\%$ in contrast to the microstructure with solute-lean matrix phase (Figure 6) observed in its counterpart with $\epsilon^{oo} = 0.1\%$ (Figure 7(e)). Furthermore, we observe the microstructures with solute-rich matrix phase (Figure 7(c)-(f)) having the highest discreteness (Figure 7(i)-(l)) in the continuously cooled alloy with higher lattice misfit $\epsilon^{oo} = 0.15\%$ on this account. For the case with $R_{elas} = 1$, both the coexisting phases tend to be more aligned along the elastically soft directions in systems with higher $\epsilon^{oo}$ and thus we observe elongated solute-rich particles in the microstructures of the isothermally aged alloys with $R_{elas} = 1$ and $\epsilon^{oo} = 0.15\%$. Also, we observed matrix-phase inversion in *AC3* with $R_{elas} = 1$ (Figure 6 and Figure 7(e)) by changing the heat treatment schedule from isothermal aging to continuous cooling only in system with $\epsilon^{oo} = 0.1\%$. For *AC3* with $R_{elas} = 1$ and $\epsilon^{oo} = 0.15\%$ (Figure 6 and Figure 7(f)), however, continuous cooling only leads to bi-continuous microstructures. This is because the solute-lean phase, which is a minor phase at high temperatures, is elongated in *AC3* with $\epsilon^{oo} = 0.15\%$ and this relatively high degree of connectivity leads to the bi-continuous microstructure instead of a microstructure with a solute-rich phase matrix as observed for the case with $\epsilon^{oo} = 0.1\%$. *This implies that the lattice misfit between the coexisting phases is an important factor that can influence the dependence of microstructual topology inversion and microstructures' discreteness on heat treatment schedule and thus a critical parameter in the design of microstructurally engineered alloys.*

### 4.5. Design of hierarchical microstructures

We have demonstrated various hierarchical microstructures (including ultrafine secondary precipitates in both the primary phase domains or in just one of the primary phase domains for a rich variety primary phase domain morphologies) that could be engineered by exploiting the miscibility gaps in systems undergoing the spinodal assisted phase transformation pathways (PTPs) via designing two-step isothermal aging heat treatments. Apart from the volume fraction of, lattice misfit, and modulus mismatch between coexisting phases, the hierarchical architecture in the two-phase microstructures evolved through such heat treatments is also strongly dependent on the asymmetry of the miscibility gap as it would determine if only one or both of the co-existing primary phases formed at the first aging temperature are inside the spinodal during the second isothermal aging step. Also, the extent of asymmetry of the miscibility gap dictates the evolution in volume fraction of the co-existing phases with temperature and thus decides the spatial arrangement and topology of the phases at the end of the first isothermal aging step, which ultimately influences the hierarchical architecture developed in the final microstructure. Further, the aging time spent during the first isothermal aging temperature dictates the scale of the primary phase domains and thus needs to be optimized to reap the desired combination of properties for targeted applications.

### 4.6. Alloy design

The connectivity and discreteness parameters of the simulated microstructures shown in Figure 7 along with the solute-rich phase fraction variation with temperature (Figure 11(a)) can serve as a microstructure map in the design of heat treatments to develop alloys with desired microstructures. Here we apply the knowledge gained from the simulated microstructures of alloys *AC1, AC2* and *AC3* (Figure 1(a)) to phase diagrams from CALPHAD databases and thereby provide three different alloys, namely $Ta_{60}Zr_{40}$, $Hf_6Ta_{57}Zr_{37}$ and $Ti_5Ta_{56}Zr_{39}$, for which the isothermal aging vs continuous cooling could potentially affect the morphology and topology of the coexisting phases in their microstructures evolved through spinodal decomposition. According to the PanNi database [48] in Pandat$^{TM}$ [37], the BCC solid solution phase in these alloys decomposes into Zr-rich and Zr-lean phases and the evolution in the volume fraction of the Zr-rich phase as a function of temperature is shown in Figure 11(b). It is readily seen that the Zr-rich phase evolves from being the major phase to the minor phase with decreasing temperature, similar to the volume fraction evolution in the prototype alloy *AC3* with $c = 0.39$ as shown in Figure 11(a) and, thus, the heat treatment schedules demonstrated in this study, i.e., isothermal aging vs continuous cooling, could potentially affect the microstructural morphology and topology. Note that the actual variation in microstructural morphology and topology is highly dependent on the modulus mismatch and lattice misfit between the coexisting Zr-rich and Zr-lean phases. Also, our CALPHAD analysis of the binary Nb-Zr, Ta-Zr and ternary Nb-Ta-Zr phase diagrams suggest that two-step isothermal aging heat treatments could be designed for alloys in these systems to engineer various hierarchical microstructures as shown in Figures 9 and 10, and experiments are underway to validate our predictions.

Although we designed the prototype binary system in our study with an asymmetric miscibility gap imitating the miscibility gap in $Al_5(TiZr)_x(NbTa)_{(95-x)}$ pseudo-binary system predicted using the RHEA database, we did not further explore this system experimentally to select potential alloys since the RHEA databases need further optimization with respect to the stability of ordered phases such as B2.

## 5. Conclusion

We investigated how heat treatment schedule (i.e., temperature-time schedule) alters the interplay among the volume fraction of, lattice misfit and modulus mismatch between coexisting phases during microstructural evolution in systems undergoing spinodal decomposition. The major findings, as summarized below, could assist in microstructure engineering of particular alloys for targeted applications.

- For systems with an asymmetric miscibility gap, the continuous matrix phase and discrete precipitate phase could be inverted for a given alloy simply by changing the heat treatment schedule from isothermal aging to continuous cooling. This matrix phase inversion is due to the temperature-dependence of the volume fractions of the co-existing phases, i.e., the evolution of a given phase from being the major phase to being the minor phase (or the other way around) with decreasing temperature. Thus, in addition to alloy composition, the heat treatment schedule design could be employed as another degree of freedom to design a rich variety of microstructures such as ordered precipitates + disordered matrix, disordered precipitates + ordered matrix, and bi-continuous mixture of ordered and disordered phases in alloys undergoing spinodal assisted phase transformation pathways (PTPs).
- Various hierarchical microstructures with ultrafine secondary precipitates embedded in one or both of the coexisting primary phase domains could be engineered in systems with symmetric or asymmetric miscibility gaps through the design of two-step isothermal aging heat treatments. The primary phases in the hierarchical microstructures developed at the end of the first isothermal aging heat treatment undergo spinodal decomposition again during the second isothermal aging step and the compositional modulations grow parallel to the pre-existing inter-phase interfaces. Thus, the spatial distribution of phases at the end of the first isothermal aging dictates the hierarchical architecture in the final microstructure. The asymmetry in the miscibility gap dictates if one or both the co-existing primary phases would further undergo spinodal decomposition and thus influences the hierarchical architecture in the two-step aged alloys. These novel hierarchical microstructures are anticipated to have some unprecedented properties.
- Our CALPHAD analysis suggests that the processing of $Ta_{60}Zr_{40}$, $Hf_6Ta_{57}Zr_{37}$ and $Ti_5Ta_{56}Zr_{39}$ through continuous cooling vs single-step isothermal aging may potentially affect the microstructure topology developed during spinodal decomposition of the BCC solid solution phase. Also, we predict that hierarchical microstructures could be engineered in alloys of Nb-Zr, Ta-Zr and Nb-Ta-Zr systems through the design of two-step isothermal aging heat treatments.
- The spinodal assisted PTPs involved in the evolution of BCC+B2 inverted superalloy like microstructure of $Al_{0.5}NbTa_{0.8}Ti_{1.5}V_{0.2}Zr$ [14,15] is an outcome of the Nb-Ta-Zr miscibility gap and thus through a careful design of appropriate alloy composition and heat treatment with the assistance of reliable CALPHAD databases and our systematic study may aid in the design of alloys with superalloy like as well as hierarchical microstructures.

**Acknowledgements:** The authors would like to acknowledge the financial support by the Army Research Laboratory and was accomplished under Cooperative Agreement Number W911NF-22-2-0032. The views and conclusions contained in this document are those of the authors and should not be interpreted as representing the official policies, either expressed or implied, of the Army Research Office or the US Government. The US Government is authorized to reproduce and distribute reprints for Government purposes notwithstanding any copyright notation herein. Also acknowledged is the support of the Air Force Office of Scientific Research (AFOSR) under grant FA9550-20-1-0015. Computational resources for this work are provided by Ohio Supercomputer Center under Project No. PAS0971.

**Tables**

Table 1. Model parameters utilized in the phase-field simulations.

| Parameter name | Symbol | Value |
|---|---|---|
| Chemical Mobility | M | 1 |
| Gradient energy coefficient of $c(\vec{r},t)$ | $\kappa_c$ | 1.00 |
| Elastic modulus tensor | $C^b_{1111} = C^b_{11}$ | 40,000 |
| | $C^b_{1122} = C^b_{12}$ | 30,000 |
| | $C^b_{2323} = C^b_{44}$ | 10,000 |
| Lattice misfit | $\epsilon^{oo}$ | Asymmetric system: 0.1%, 0.15% <br> Symmetric system: 0.05%, 0.1% |
| Modulus mismatch | $R_{elas} = \dfrac{C^y_{ijkl}}{C^b_{ijkl}}$ | 1.0 (Homo) <br> 1.4 (Hard yellow-phase) <br> 0.6 (Soft yellow-phase) |
| Discretization timestep in single-step isothermal aging and continuous cooling heat treatments | $\Delta t$ | 0.01 |
| Total duration of the simulation in single-step isothermal aging and continuous cooling heat treatments | $T^{total}$ | 28000 |

**Figures**

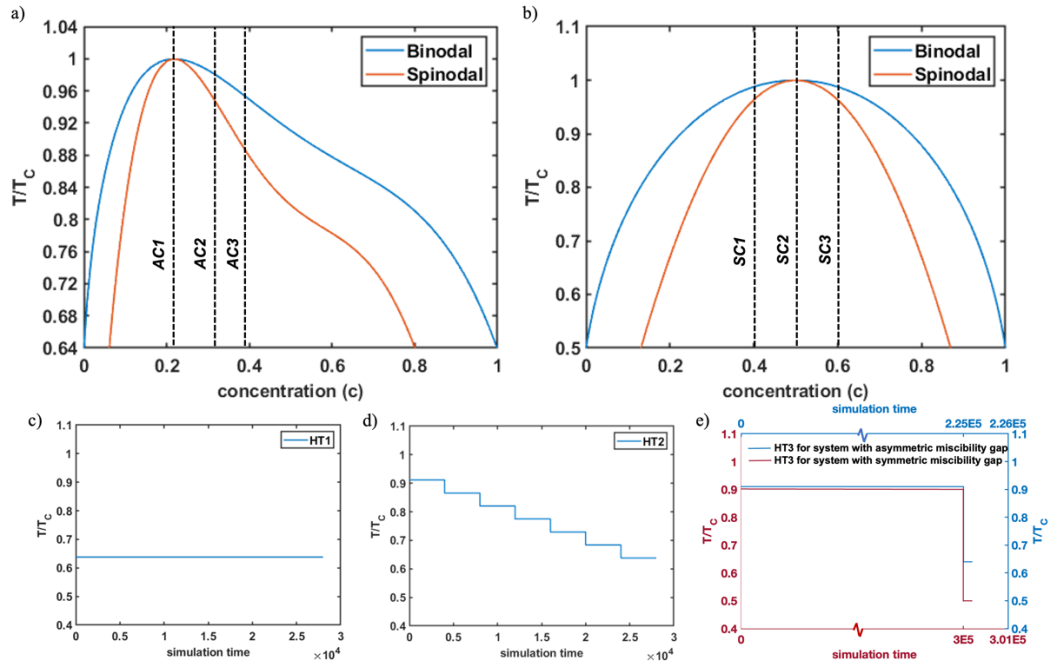

Figure 1. a) Normalized phase diagram of the prototype binary system with asymmetric miscibility gap. b) Normalized phase diagram of the prototype binary system with symmetric miscibility gap. c) One-step isothermal aging schedule (HT1). d) continuous cooling schedule (HT2). e) Two-step isothermal aging schedule (HT3).

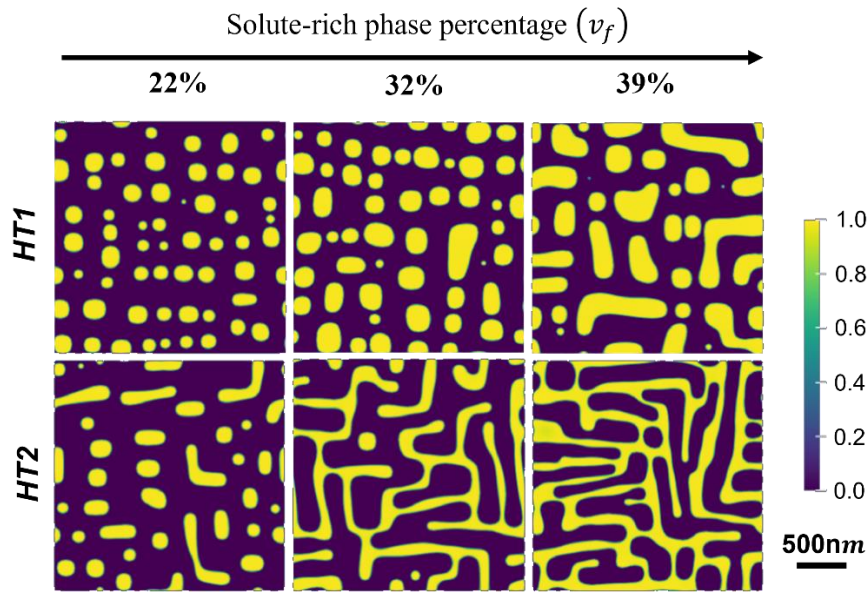

Figure 2. Effect of single-step isothermal aging (HT1) vs continuous cooling (HT2) on the microstructure topology ($t = 28000$) in alloys with different $v_f$ under homogeneous modulus condition. The color contrast in the microstructures represent the variation in solute concentration ($c$).

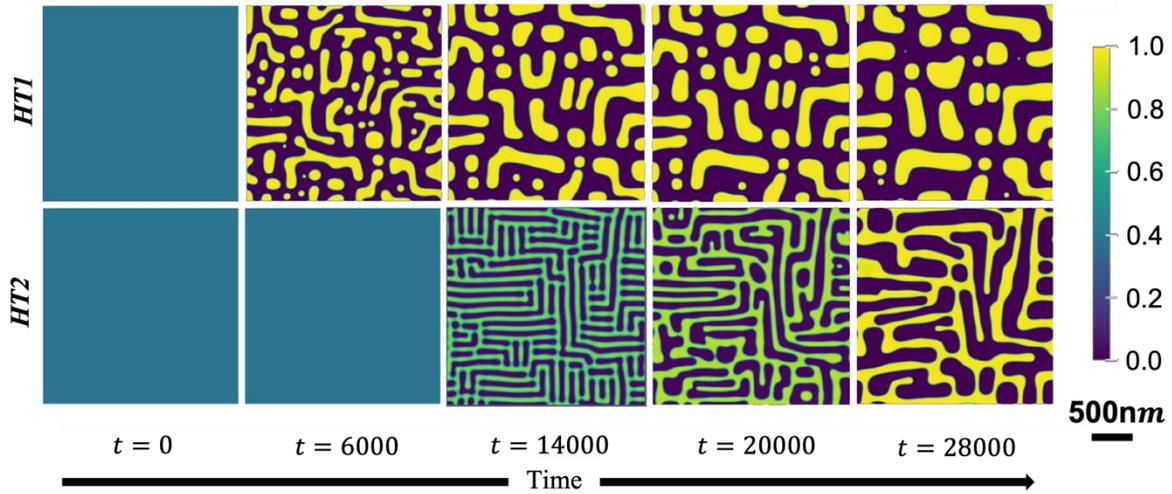

Figure 3. Microstructure evolution in *asymmetric alloys with c=0.39* processed through HT1 and HT2 with homogeneous modulus and lattice misfit $\epsilon^{oo} = 0.1\%$.

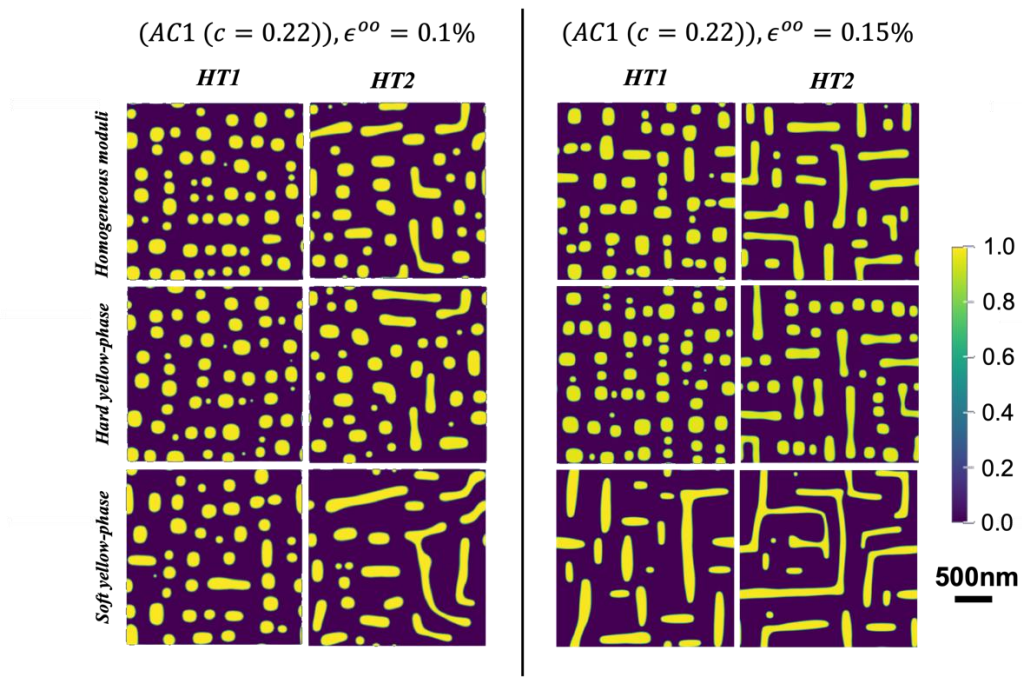

Figure 4. Effect of lattice misfit ($\epsilon^{oo}$) and moduli mismatch ($R_{elas}$) on the microstructure ($t = 28000$) of asymmetric prototype alloy $AC1$ ($c = 0.22$) processed through one-step isothermal aging (HT1) and continuous cooling (HT2) heat treatments.

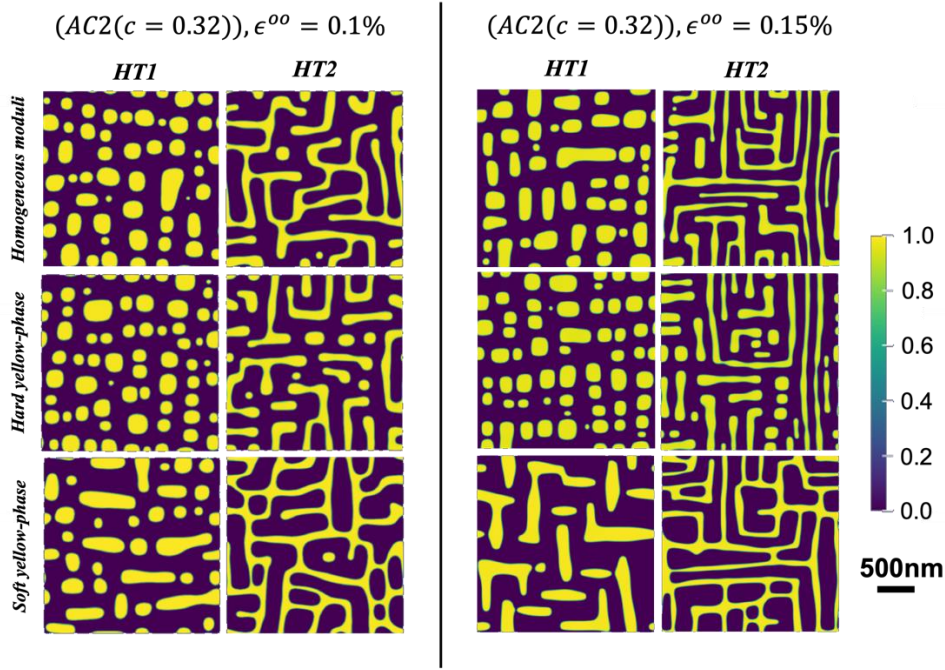

Figure 5. Effect of lattice misfit ($\epsilon^{oo}$) and moduli mismatch ($R_{elas}$) on microstructure ($t = 28000$) of asymmetric prototype alloy $AC2$ ($c = 0.32$) processed through one-step isothermal aging (HT1) and continuous cooling (HT2) heat treatments.

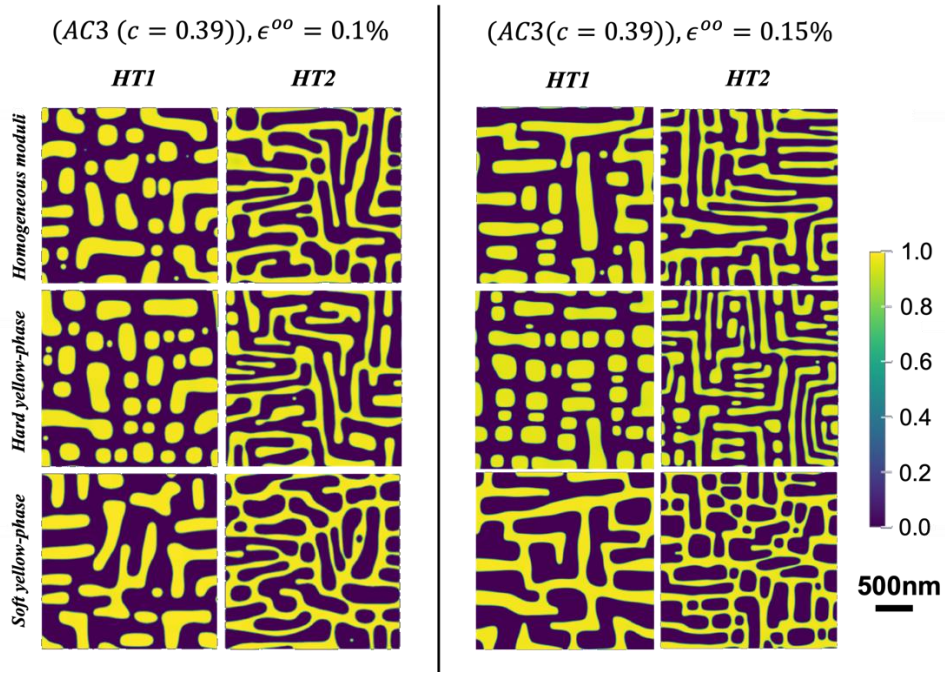

Figure 6. Effect of lattice misfit ($\epsilon^{oo}$) and moduli mismatch ($R_{elas}$) on microstructure ($t = 28000$) of asymmetric prototype alloy $AC3$ ($c = 0.39$) processed through one-step isothermal aging (HT1) and continuous cooling (HT2) heat treatments.

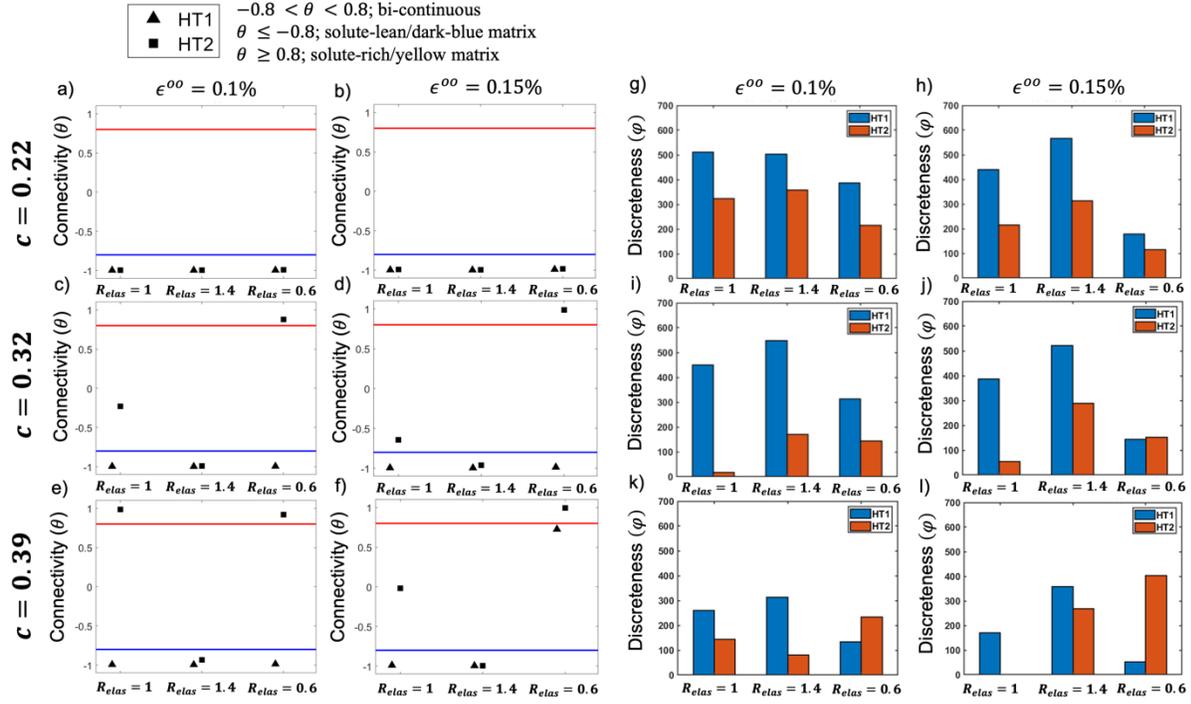

Figure 7. Quantification of the connectivity ($\theta$) and discreteness ($\varphi$) for the simulated microstructures of alloys with asymmetric miscibility gap.

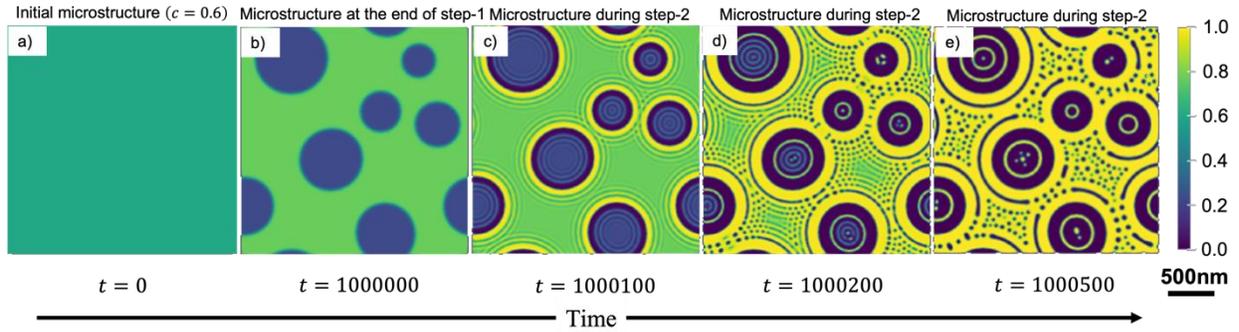

Figure 8. Microstructure evolution in symmetric alloy *SC1* ($c = 0.6$) processed through two-step isothermal aging heat treatment (HT3) under the assumption of zero lattice misfit.

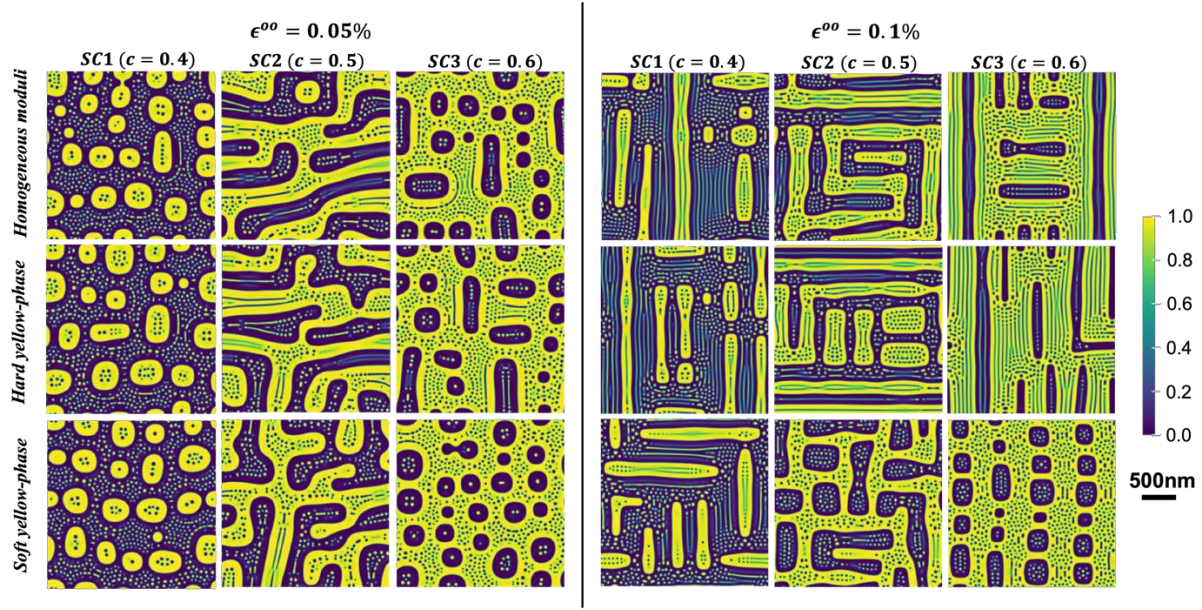

Figure 9. Simulated microstructures of two-step isothermally aged symmetric alloys ($t = 300200$).

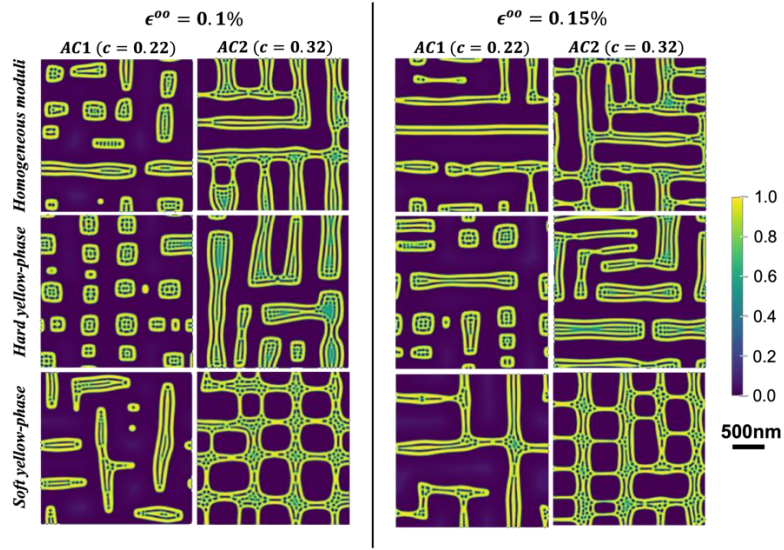

Figure 10. Simulated microstructures of two-step isothermally aged asymmetric alloys ($t = 225200$).

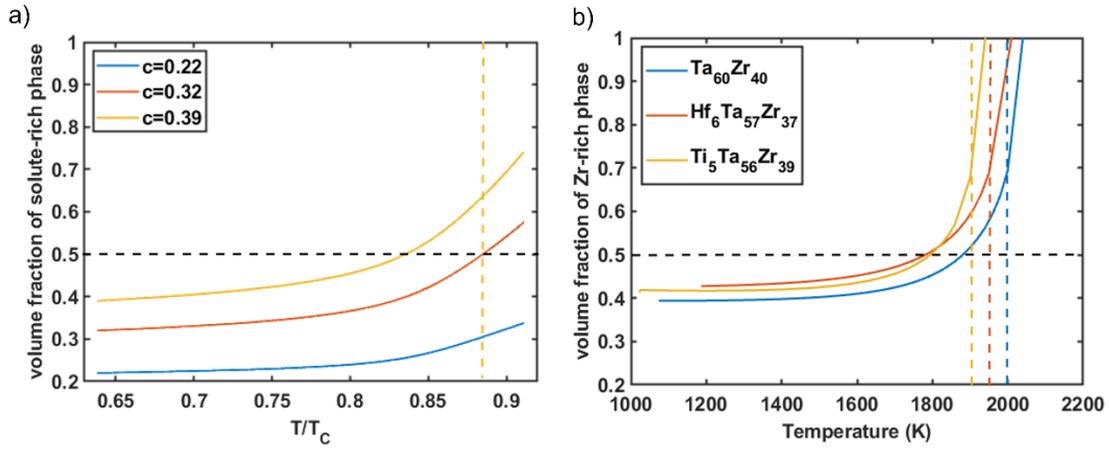

Figure 11. a) Solute-rich phase fraction vs normalized temperature for prototype alloys with asymmetric miscibility gap; b) Volume fraction of Zr-rich phase vs temperature for the alloys $Ta_{60}Zr_{40}$, $Hf_6Ta_{57}Zr_{37}$ and $Ti_5Ta_{56}Zr_{39}$ calculated with PanNi database [48]. The dotted lines indicate the spinodal temperature. Spinodal temperatures for the prototype alloys $c = 0.22$ and $c = 0.32$ are missing in (a) as they are higher than $0.91 T_c$.